\documentclass[12pt]{article}
\usepackage{graphicx}

\def\a{\alpha}
\def\b{\beta}

\def\e{\epsilon}

\def\g{\gamma}

\def\m{\mu}
\def\n{\nu}

\def\p{\pi}

\def\s{\sigma}
\def\t{\tau}

\def\D{\Delta}

\def\G{\Gamma}

\def\beq{\begin{equation}}
\def\eeq{\end{equation}}
\def\bea{\begin{eqnarray}}
\def\eea{\end{eqnarray}}

\def\pl#1#2#3{Phys.~Lett.~{\bf B {#1}} (19{#2}) #3}
\def\np#1#2#3{Nucl.~Phys.~{\bf B {#1}} (19{#2}) #3}
\def\prl#1#2#3{Phys.~Rev.~Lett.~{\bf #1} (19{#2}) #3}
\def\pr#1#2#3{Phys.~Rev.~{\bf D {#1}} (19{#2}) #3}

\def\prep#1#2#3{Phys.~Rep.~{\bf {#1}C} (19{#2}) #3}

\newcommand{\gl}{\tilde{g}}

\newcommand{\wi}{\tilde{w}}
\newcommand{\bi}{\tilde{b}}

\begin{document}
\date{\mbox{ }}

\title{ 
{\normalsize     
DESY 99-169\hfill\mbox{}\\
December 1999\hfill\mbox{}\\}
\vspace{2cm}
{\bf NEUTRINO MIXING AND\\ FLAVOUR CHANGING PROCESSES}\\
%
\author{{Wilfried~Buchm\"uller}\thanks{E-mail : buchmuwi@vxdesy.desy.de}, 
        {David~Delepine}\thanks{E-mail : delepine@mail.desy.de}, 
        {Laksana~Tri~Handoko}\thanks{E-mail : handoko@mail.desy.de} \\
        \vspace{1mm}
        {\it Deutsches Elektronen-Synchrotron DESY, Hamburg, Germany}}}

\maketitle

\thispagestyle{empty}

\begin{abstract}
\noindent We study the implications of a large $\nu_\m$-$\nu_\t$ mixing
angle on flavour changing transitions of quarks and leptons in supersymmetric 
extensions of the standard model. Two patterns of supersymmetry breaking
are considered, models with modular invariance and the standard scenario
of universal soft breaking terms at the GUT scale. The analysis is performed
for two symmetry groups $G\otimes U(1)_F$, with $G=SU(5)$ and $G=SU(3)^3$, 
where $U(1)_F$ is a family symmetry. Models with modular invariance are
in agreement with observations only for restricted scalar quark and gaugino
masses, $\overline{M}_q^2/m_{\gl}^2 \simeq 7/9$ and $m_{\bi} > 350$~GeV. 
A characteristic feature of models with large $\tan{\b}$ and radiatively 
induced flavour mixing is a large branching ratio for $\m \rightarrow e\g$. 
For both symmetry groups and for the considered range of supersymmetry
breaking mass parameters we find $\mathrm{BR}(\m \rightarrow e\g)>10^{-14}$.
\end{abstract}

\newpage

\section{Introduction}

The recently reported atmospheric neutrino anomaly \cite{atm98} can be
interpreted as a manifestation of neutrino oscillations with a large
$\n_\m - \n_\t$ mixing angle. The smallness of the corresponding neutrino
masses is naturally explained by the seesaw mechanism \cite{yan79}, which
leads to the prediction of heavy Majorana neutrinos with masses close to 
the unification scale $\Lambda_{GUT}$.

A $\n_\m - \n_\t$ mixing angle, which is large compared to the Cabbibo angle,
requires an explanation within a unified theory of leptons and quarks.  
An attractive class of models for lepton and quark mass matrices is based
on the symmetries $G\otimes U(1)_F$, where $G$ is a unified gauge group
and $U(1)_F$ is a family symmetry \cite{fn79}-\cite{lr99}. The large neutrino 
mixing angle can then be explained  either by a non-parallel family 
structure of chiral charges \cite{sy98}, or by a parametrically large 
flavour mixing \cite{lr99}. Both possibilites are phenomenologically
viable and can also account for the cosmological baryon asymmetry by
means of heavy Majorana neutrino decays \cite{bp99}.

The large hierarchy between the electroweak scale and the unification scale,
and now also the mass scale of the heavy Majorana neutrinos, motivates
supersymmetric extensions of the standard model \cite{nil84}. This is
further supported by the observed unification of gauge couplings. The least
understood aspect of the supersymmetric standard model is the mechanism of
supersymmetry breaking and the corresponding structure of soft supersymmetry
breaking masses and couplings.

Constraints from rare processes severely restrict the allowed pattern of 
supersymmetry breaking \cite{en82}-\cite{mpr98}. In the standard scenario 
with universal soft breaking terms at the GUT scale radiative corrections 
induce flavour mixing at the electroweak scale. Alternatively, an interesting 
class of models with modular invariance \cite{il92} predicts non-universal
soft breaking terms at the GUT scale at tree level. In models with a
$U(1)_F$ family symmetry the structure of these mass matrices is
determined by the chiral charges of quarks and leptons \cite{dps96}.

Flavour changing hadronic processes have been studied for models with radiative
flavour mixing as well as for models with modular invariance 
\cite{mpr98,dps96}. Particularly interesting processes 
are lepton flavour changing radiative transitions \cite{bh94}-\cite{ko99}. 
Here large Yukawa couplings or flavour changing tree level scalar mass terms 
can lead to predictions for branching ratios comparable to the present 
experimental limits. The dependence on the underlying flavour symmetry
has been studied in \cite{lt98}-\cite{egl99}. 

In this paper we extend the analysis of \cite{bdv99}. We shall compare two 
symmetry groups, $SU(5)\otimes U(1)_F$ and $SU(3)^3\otimes U(1)_F$, both
with radiatively induced flavour mixing and with modular invariance,
respectively. In Sec.~2 we present Yukawa couplings and scalar mass
matrices. Sec.~3 deals with quark flavour changing processes, and in
Sec.~4 flavour changing radiative transitions are discussed, leading
to the conlusions in Sec.~5.

\section{Patterns of supersymmetry breaking}

\subsection{Yukawa couplings and scalar masses}

We consider the supersymmetric standard model with right-handed neutrinos,
which is described by the superpotential 
\begin{eqnarray}
  W & = & \mu H_1 H_2 + {h_e}_{ij} E_i^c L_j H_1 + 
  {h_\nu}_{ij} N_i^c L_j H_2 + 
  \frac{1}{2} {h_r}_{ij} N_i^c N_j^c R  
  \nonumber \\
  & & + {h_d}_{ij} Q_i D_j^c H_1 + 
  {h_u}_{ij} Q_i U_j^c H_2 \; .
\end{eqnarray}
Here $i,j=1\ldots 3$ are generation indices; the superfields $E^c$, $L=(N,E)$%
, $N^c$ contain the leptons $e_R^c$, $(\nu_L,e_L)$, $\nu_R^c$, respectively,
and the superfields $U^c$, $Q=(U,D)$, $D^c$ contain the quarks $u_R^c$, $%
(u_L,d_L)$, $d_R^c$. The expectation values of the Higgs multiplets $H_1$
and $H_2$ generate ordinary Dirac masses of quarks and leptons, and the
expectation value of the singlet Higgs field $R$ yields the Majorana mass
matrix of the right-handed neutrinos.

In the following discussion the scalar masses will play a crucial role. They
are determined by the superpotential and the soft breaking terms, 
\begin{eqnarray}
  L_{\rm soft} & = & - (\widetilde{m}_l^2)_{ij} L_i^\dagger L_j -
  (\widetilde{m}_e^2)_{ij} E_i^{c^\dagger} E_j^c - 
  (\widetilde{m}_q^2)_{ij} Q_i^\dagger Q_j -
  (\widetilde{m}_d^2)_{ij} D_i^{c^\dagger} D_j^c -
  (\widetilde{m}_u^2)_{ij} U_i^{c^\dagger} U_j^c  
  \nonumber \\
  & & + {A_e}_{ij} E_i^c L_j H_1 + 
  {A_\nu}_{ij} N_i^c L_j H_2 + 
  {A_d}_{ij} Q_i D_j^c H_1 + 
  {A_u}_{ij} Q_i U_j^c H_2 + 
  c.c. + \cdots \; , 
  \label{soft}
\end{eqnarray}
where $L=(N_L,E_L)$, $E^c = E_R^*$, $Q=(U_L,D_L)$, $D^c = D_R^*$,
and $U^c = U_R^*$ denote the scalar partners of $(\nu_L, e_L)$, $e_R^c$%
, $(u_L,d_L)$, $d_R^c$ and $u_R^c$, respectively. Using the seesaw mechanism
to explain the smallness of neutrino masses, we assume that the right-handed
neutrino masses $M$ are much larger than the Fermi scale $v$. One then
easily verifies that all mixing effects on light scalar masses caused by the
right-handed neutrinos and their scalar partners are suppressed by $\mathcal{%
O}(v/M)$, and therefore negligable.

The scalar mass terms are then given by 
\begin{equation}
L_{M}=-E^{\dagger }{\widetilde{M}_{e}}^{2}E-N_{L}^{\dagger }{\widetilde{m}%
_{l}}^{2}N_{L}\;-D^{\dagger }{\widetilde{M}_{d}}^{2}D-U^{\dagger }{%
\widetilde{M}_{u}}^{2}U\;,
\end{equation}
where $\widetilde{M}_{e}^{2}$ is the mass matrix of the scalar fields $%
E=(E_{L},E_{R})$, 
\begin{equation}
{\widetilde{M}_{e}}^{2}\ \equiv \ \left( 
\begin{array}{ll}
\widetilde{M}_{eL}^{2} & \widetilde{M}_{eLR}^{2} \\ 
\widetilde{M}_{eRL}^{2} & \widetilde{M}_{eR}^{2}
\end{array}
\right) \ =\ \;\left( 
\begin{array}{ll}
{\widetilde{m}_{l}}^{2}+v_{1}^{2}h_{e}^{\dagger }h_{e} & v_{1}A_{e}^{\dagger
}+\mu v_{2}h_{e}^{\dagger } \\[1ex]
v_{1}A_{e}+\mu v_{2}h_{e} & {\widetilde{m}_{e}}^{2\dagger}+
v_{1}^{2}h_{e}h_{e}^{\dagger }
\end{array}
\right) \;,  \label{escal}
\end{equation}
$\widetilde{M}_{d}^{2}$ is the mass matrix of the scalar fields $%
D=(D_{L},D_{R})$, 
\begin{equation}
{\widetilde{M}_{d}}^{2}\ \equiv \ \left( 
\begin{array}{ll}
\widetilde{M}_{dL}^2 & \widetilde{M}_{dLR}^{2} \\ 
\widetilde{M}_{dRL}^{2} & \widetilde{M}_{dR}^{2}
\end{array}
\right) \ =\ \;\left( 
\begin{array}{ll}
{\widetilde{m}_{q}}^{2}+v_{1}^{2}h_{d}^{*}h_{d}^{T} & v_{1}A_{d}^{*}+\mu
v_{2}h_{d}^{*} \\[1ex]
v_{1}A_{d}^{T}+\mu v_{2}h_{d}^{T} & {\widetilde{m}_{d}}^{2\dagger}+
v_{1}^{2}h_{d}^{T}h_{d}^{*}
\end{array}
\right) \;,
\end{equation}
and $\widetilde{M}_{u}^{2}$ is the mass matrix of the scalar fields $%
U=(U_{L},U_{R})$, 
\begin{equation}
{\widetilde{M}_{u}}^{2}\ \equiv \ \left( 
\begin{array}{ll}
\widetilde{M}_{uL}^2 & \widetilde{M}_{uLR}^{2} \\ 
\widetilde{M}_{uRL}^{2} & \widetilde{M}_{uR}^{2}
\end{array}
\right) \ =\ \left( 
\begin{array}{ll}
{\widetilde{m}_{q}}^{2}+v_{2}^{2}h_{u}^{*}h_{u}^{T} & v_{2}A_{u}^{*}+\mu
v_{1}h_{u}^{*} \\[1ex]
v_{2}A_{u}^{T}+\mu v_{1}h_{u}^{T} & {\widetilde{m}_{u}}^{2\dagger}+
v_{2}^{2}h_{u}^{T}h_{u}^{*}
\end{array}
\right) \;.
\end{equation}

According to the Froggatt-Nielsen mechanism \cite{fn79} the hierarchies
among the various Yukawa couplings are related to a spontaneously broken U(1)%
$_{F}$ generation symmetry. The Yukawa couplings arise from
non-renormalizable interactions after a gauge singlet field $\phi$ aquires
a vacuum expectation value, 
\begin{equation}
h_{ij}=g_{ij}\left( {\frac{\left\langle \phi \right\rangle }{\Lambda }}%
\right) ^{X_{i}+X_{j}}\;.
\end{equation}
Here $g_{ij}$ are couplings $\mathcal{O}(1)$ and $X_{i}$ are the U(1)
charges of the various superfields with $X_{\phi }=-1$. The interaction
scale $\Lambda $ is expected to be very large, $\Lambda >\Lambda _{GUT}$.

For $G=SU(5)$, quarks and leptons are grouped into the multiplets $\mathbf{10%
}=(q_L,u_R^c,e_R^c)$, $\mathbf{5^*}=(d_R^c,l_L)$ and $\mathbf{1}=\nu_R^c$.
The phenomenology of quark and lepton mass matrices can be accounted for
assuming 
\begin{equation}
\left({\frac{\left\langle \phi \right\rangle }{\Lambda}}\right)^2 =
\epsilon^2 \simeq {\frac{1}{300}}\;.
\end{equation}
The corresponding $U(1)_F$ charges are given in Tab.~\ref{tab:su5} 
\cite{by99}. 
The same charge assignment to the lepton doublets of the second and third
generation leads to a large $\nu_{\mu}-\nu_{\tau}$ mixing angle \cite{sy98}. 
As in all $SU(5)$ GUTs the difference between the down-quark mass
hierarchy and the charged lepton mass hierarchy has to be explained by some
additional mechanism \cite{gj79}-\cite{ikn99}. 
\begin{table}[h]
\begin{center}
\begin{tabular}{c|ccccccccc}
\hline\hline
$\psi _{i}$ & $\mathbf{10}_3$ & $\mathbf{10}_2$ & $\mathbf{10}_1$ & $\mathbf{%
5^*}_3$ & $\mathbf{5^*}_2$ & $\mathbf{5^*}_1$ & $\mathbf{1}_3$ & $\mathbf{1}%
_2 $ & $\mathbf{1}_1$ \\ \hline
$X_{i}$ & $0$ & $1$ & $2$ & $a$ & $a$ & $a+1$ & $0$ & $1-a$ & $2-a$ \\ 
\hline\hline
\end{tabular}
\caption[dum]{\textit{$U(1)_F$ charges for quarks and leptons; $G=SU(5)$,
a=0 or 1.}}
\label{tab:su5}
\end{center}
\end{table}
In the following we choose $a=0$, i.e. the case of large down-quark and
charged lepton Yukawa couplings which corresponds to $\tan{\b}\sim 1/\e$.
The case $a=1$ leads to significantly smaller rates for flavour changing
processes. For lepton flavour changing radiative transitions this case has been
discussed in \cite{bdv99}.
 
The Yukawa matrices corresponding to the charges in Tab.~\ref{tab:su5}  
have the structure 
\begin{equation}
h_{d}\sim h_{e}\sim h_{\n}\sim \left( 
\begin{array}{lll}
\epsilon ^{3} & \epsilon ^{2} & \epsilon ^{2} \\ 
\epsilon ^{2} & \epsilon  & \epsilon  \\ 
\epsilon  & 1 & 1
\end{array}
\right) \;,\quad h_{u}\sim \left( 
\begin{array}{lll}
\epsilon ^{4} & \epsilon ^{3} & \epsilon ^{2} \\ 
\epsilon ^{3} & \epsilon ^{2} & \epsilon  \\ 
\epsilon ^{2} & \epsilon  & 1
\end{array}
\right) \;.
\end{equation}
The difference between the Yukawa matrices $h_{u}$ and $h_{d}$ yields the
CKM matrix, 
\begin{equation}
V_{CKM}\sim \left( 
\begin{array}{lll}
1 & \epsilon  & \epsilon ^{2} \\ 
\epsilon  & 1 & \epsilon  \\ 
\epsilon ^{2} & \epsilon  & 1
\end{array}
\right) \;,  \label{ckm5}
\end{equation}
which is very close to the measured CKM matrix. The largest factors $%
\mathcal{O}(1)$ are needed for $V_{us}$ and $V_{cd}$, since
$(V_{us})_{\exp }\simeq 4\epsilon $.

For $G=SU(3)^{3}$, quarks and leptons are assigned to different multiplets: 
${\bf L}=(l_{L},e_{R}^{c},\nu _{R}^{c},\ldots )$, 
${\bf Q_{L}}=(q_{L},\ldots )$ and ${\bf Q_{R}}=(u_{R}^{c},d_{R}^{c},\ldots )$.
A successful description of lepton and
quark masses and mixings can be achieved with the charge assignment given in
Tab. \ref{tab:u1f} 
\cite{lr99}. Contrary to the $SU(5)$ model different mass scales $%
\Lambda _{1}$ and $\Lambda _{2}$ are assumed for the Yukawa couplings of the
Higgs fields $H_{1}$ and $H_{2}$, respectively, 
\beq
h_{dij}=g_{dij}\left( \frac{\left\langle \phi \right\rangle }{\Lambda_1}%
\right) ^{X_{i}+X_{j}}\;,\quad 
h_{eij}=g_{eij}\left( \frac{\left\langle \phi
\right\rangle }{\Lambda _1}\right) ^{X_{i}+X_{j}}\;,
\eeq
\beq
h_{uij}=g_{uij}\left( \frac{\left\langle \phi \right\rangle }{\Lambda _2}%
\right) ^{X_{i}+X_{j}}\;,\quad 
h_{\nu ij}=g_{\nu ij}\left( \frac{%
\left\langle \phi \right\rangle }{\Lambda_2}\right) ^{X_{i}+X_{j}}\;,
\eeq
where all $g_{ij}$ are couplings $\mathcal{O}(1)$. The phenomenology of
quark and lepton mass matrices and a large mixing between $\nu _{\mu }$ and $%
\nu _{\tau }$ can be explained with 
\beq
\left( \frac{\left\langle \phi \right\rangle }{\Lambda _{1}}\right)
^{2}= \overline{\epsilon }^{2}\simeq \frac{1}{16}\;,\quad \left( \frac{%
\left\langle \phi \right\rangle }{\Lambda _{2}}\right) ^{2}=\overline{%
\epsilon }^{4}\simeq \epsilon ^{2}\;.  \label{epsilon}
\eeq
Note, that the effective flavor mixing $\overline{\epsilon }^{1/2}=1/2$ is much
larger than the flavor mixing $\epsilon =1/17$ in the $SU(5)$ model.

\begin{table}[h]
\begin{center}
\begin{tabular}{c|ccccccccc}
\hline\hline
$\psi_i $ & ${\bf L_3}$ & ${\bf L_2}$ & ${\bf L_1}$ & ${\bf Q_{L3}}$ & 
${\bf Q_{L2}} $ & ${\bf Q_{L1}}$ & ${\bf Q_{R3}} $ & ${\bf Q_{R2}}$ & 
${\bf Q_{R1}}$ \\ 
\hline
$X_{i}$ & $0$ & ${\frac{1}{2}}$ & ${\frac{5}{2}}$ & $0 $ & $2 $ & $3 $ & $0 $
& $0 $ & $1 $ \\ 
\hline\hline
\end{tabular}
\caption[dum]{\textit{$U(1)_F$ charges for quarks and leptons; 
    $G=SU(3)^3$.}}
\label{tab:u1f}
\end{center}
\end{table}

From Tab.~\ref{tab:u1f} one reads off the structure of the Yukawa matrices, 
\begin{eqnarray}
h_{e} &\sim &\left( 
\begin{array}{lll}
\overline{\epsilon}^5 & \overline{\epsilon}^3 & \overline{\epsilon}^{5/2} \\ 
\overline{\epsilon}^3 & \overline{\epsilon} & \overline{\epsilon}^{1/2} \\ 
\overline{\epsilon}^{5/2} & \overline{\epsilon}^{1/2} & 1
\end{array}
\right)\;,\quad h_{\nu }\sim \left( 
\begin{array}{lll}
\epsilon^5 & \epsilon^3 & \epsilon^{5/2} \\ 
\epsilon^3 & \epsilon  & \epsilon^{1/2} \\ 
\epsilon^{5/2} & \epsilon^{1/2} & 1
\end{array}
\right) \;, \nonumber \\
h_{d} &\sim &\left( 
\begin{array}{lll}
\overline{\epsilon }^{4} & \overline{\epsilon }^{3} & \overline{\epsilon }%
^{3} \\ 
\overline{\epsilon }^{3} & \overline{\epsilon }^{2} & \overline{\epsilon }%
^{2} \\ 
\overline{\epsilon } & 1 & 1
\end{array}
\right) \;,\quad h_{u}\sim \left( 
\begin{array}{lll}
\epsilon ^{4} & \epsilon ^{3} & \epsilon ^{3} \\ 
\epsilon ^{3} & \epsilon ^{2} & \epsilon ^{2} \\ 
\epsilon  & 1 & 1
\end{array}
\right) \;.  
\end{eqnarray}
The corresponding CKM matrix is given by
\begin{equation}
V_{CKM}\sim \left( 
\begin{array}{lll}
1 & \epsilon ^{1/2} & \epsilon ^{3/2} \\ 
\epsilon ^{1/2} & 1 & \epsilon  \\ 
\epsilon ^{3/2} & \epsilon  & 1
\end{array}
\right) \;,  \label{ckm3}
\end{equation}
which is also very close to the measured CKM matrix. In this case the
smallest factors $\mathcal{O}(1)$ are needed for $V_{ub}$ and $V_{td}$,
since $(V_{ub})_{\exp }\simeq \frac{1}{4} \epsilon^{3/2}$.

\subsection{ Soft breaking terms from modular invariance}

For a wide class of supergravity models the possibilities of supersymmetry
breaking can be parametrized by vacuum expectation values of moduli fields $%
T_a$ and the dilaton field $S$ \cite{bim94}. The structure of the soft
breaking terms is determined by the modular weights of the various
superfields. An interesting structure arises if the theory possesses both,
modular invariance and a chiral $U(1)$ symmetry. Under the modular
transformation, the moduli fields $T_{a}$ and the matter field $\Phi$
transform like 
\begin{eqnarray}
  T_{a} &\rightarrow &(a_{a}T^{a}-ib_{a})/(ic_{a}T^{a}+d_{a}) \; ,  
  \nonumber \\
  \Phi ^{i} &\rightarrow &(ic_{a}T^{a}+d_{a})^{n_{i}^{(a)}} \Phi ^{i} \;,
\end{eqnarray}
with $a_{a}d_{a}-b_{a}c_{a}=1$ and $a_{a},b_{a},c_{a},d_{a}\in Z$. Here $%
n_i^{(a)}$ is the modular weight of the field $\Phi ^{i}$ with respect to
the moduli field $T_a$. Consider now superpotential and K\"ahler potential
for quark fields, moduli fields and dilaton, 
\begin{eqnarray}
  W & = & {h_d}_{ij} \theta (X_{i} + X_j) Q_i D_j^c H_1 
  \left( \frac{\phi }{\Lambda _{1}}\right) ^{X_{i}+X_{j}} +h_{uij}\theta
  (X_{i}+X_{j})Q_{i}U_{j}^{c}H_{2}\left( \frac{\phi }{\Lambda _{2}}
  \right)^{X_{i}+X_{j}}\;, \\
  K & = & K_{0}(T^{a},\bar{T}^{a})-\ln (S+\bar{S}) + 
  \prod_{a} t_{a}^{n_\Phi^{(a)}} \bar{\Phi} \Phi +
  K_{\bar{i}j}\bar{\Phi}^{\bar{i}} \Phi^j \; , \\
  K_{\bar{i}j} &=&\delta _{ij}\prod_{a}t_{a}^{n_{j}^{(a)}} + 
  Z_{\bar{i}j} \left[ \theta (X_{i} - X_{j})\prod_{a}t_{a}^{n_{j}^{(a)}} 
    \left( \frac{\bar{\phi}}{\Lambda_{3}} \right)^{X_{i}-X_{j}} \right.  
  \nonumber \\
  & & \left.
    \hspace{2.5cm}+\theta (X_{j} - X_{i})\prod_{a}t_{a}^{n_{i}^{(a)}} 
    \left( \frac{\phi }{\Lambda _{3}} 
    \right)^{X_{j}-X_{i}}\right] + \cdots \; .
\end{eqnarray}
Here $X_{i}$ are the $U(1)_{F}$ charges of the matter fields $\Phi=
Q_i, D^c_j, U^c_k$, $t_{a}= T_{a}+\bar{T}_{a}$, and $\Lambda_1$, $\Lambda_2$%
, $\Lambda_3$ are three mass scales. Under a modular transformation $K_{0}$
transforms as 
\begin{equation}
K_{0}\rightarrow K_{0}+n_{0}^{(a)}\ln \left| ic_{a}T_{a}+d_{a}\right|^{2}\;,
\end{equation}
whereas $G\equiv K+\ln \left| W\right| ^{2}$ has to be invariant. This
yields a relation between the modular weights and the $U(1)_{F}$ charges 
\cite{dps96}. One easily verifies that the invariance of $G$ holds for
arbitrary mass scales $\Lambda_1$, $\Lambda_2$ and $\Lambda_3$.

The supersymmetry breaking scalar mass terms are directly related to the
charges of the corresponding superfields \cite{dps96}, 
\beq
{\widetilde{m}}_{ij}^{2}=\left( (1+B_{i}(\Theta _a))\delta
_{ij}+|X_{i}-X_{j}|C_{ij}(\Theta_a)\widetilde{\epsilon}%
^{|X_{i}-X_{j}|} \right) M^{2}\;,  \label{modular}
\eeq
where $\widetilde{\epsilon}=(\phi/\Lambda_3)$ and the $\Theta_a$ parametrize
the direction of the goldstino in moduli space. For pure dilaton breaking, 
i.e. $\Theta_a=0$, one has $C_{ij}=0$ and the soft breaking terms are flavour
diagonal.

In the $SU(5)$ model, $\epsilon$ is the same for all Yukawa couplings, i.e. $%
\Lambda_{1} = \Lambda_{2} \equiv \Lambda$. Hence, for simplicity, we also
choose $\Lambda_{3} = \Lambda$. For the scalar lepton and quark mass
matrices one then obtains, 
\beq
{\widetilde{m}}_{d}^{2}\sim {\widetilde{m}}_{l}^{2}\sim \ \left( 
\begin{array}{ccc}
1 & \epsilon & \epsilon \\[1ex] 
\epsilon & 1 & 0 \\[1ex] 
\epsilon & 0 & 1
\end{array}
\right) \ M^{2}\;, \quad {\widetilde{m}}_{q}^{2}\sim {\widetilde{m}}%
_{u}^{2}\sim {\widetilde{m}}_{e}^{2}\sim \ \left( 
\begin{array}{ccc}
1 & \epsilon & \epsilon ^{2} \\[1ex] 
\epsilon & 1 & \epsilon \\[1ex] 
\epsilon ^{2} & \epsilon & 1
\end{array}
\right) \ M^{2}\;.  \label{softmod}
\eeq
Note, that the zeros in ${\widetilde{m}}_{l}^{2}$ occur since the lepton
doublets of the second and the third family carry the same U(1)$_{F}$ charge.

The scalar mass matrices (\ref{softmod}) are given in the weak eigenstate
basis. In order to discuss the radiative transitions $\mu \rightarrow
e\gamma $, $\tau \rightarrow \mu \gamma $, $b\rightarrow s\gamma $, and the
mixing parameters $\Delta M_K$, $\Delta M_{B_d}$, $\Delta M_{B_s}$, 
$\epsilon_K$, we have to
change to a mass eigenstate basis of charged leptons and down quarks. The
Yukawa matrix $h_{e}=h_{d}$ can be diagonalized by a bi-unitary
transformation, $U^{\dagger }h_{e}V=h_{e}^{D}$. To leading order in $\e$ 
the matrices $U$ and $V$ read 
\begin{equation}
U=\left( 
\begin{array}{ccc}
1 & a\epsilon  & b\epsilon ^{2} \\[1ex]
-a\epsilon  & 1 & f\epsilon  \\[1ex]
-b\epsilon ^{2} & -f\epsilon  & 1
\end{array}
\right) \;,\quad V=\left( 
\begin{array}{ccc}
1 & (ca^{\prime }-sb^{\prime })\epsilon  & (sa^{\prime }+cb^{\prime
})\epsilon  \\[1ex]
-a^{\prime }\epsilon  & c & s \\[1ex]
-b^{\prime }\epsilon  & -s & c
\end{array}
\right) \;,
\end{equation}
where $c=\cos {\varphi }$ and $s=\sin {\varphi }$; $a,b,a^{\prime
},c^{\prime }$ depend on the coefficients $\mathcal{O}(1)$ in the Yukawa
matrices. The scalar mass matrices transform as $V^{\dagger }\widetilde{m}%
_{l,q}^{2}V$, $U^{\dagger }\widetilde{m}_{e,d}^{2}U$. One easily verifies
that the form of the matrix $\widetilde{m}_{e}^{2}$ is invariant, whereas
the matrices $\widetilde{m}_l^2$, $\widetilde{m}_q^2$, 
$\widetilde{m}_d^2$ become after diagonalisation,
\beq\label{general}
{\widetilde{m}}_{l}^{2}\sim {\widetilde{m}}_{q}^{2}\sim \ \left( 
\begin{array}{ccc}
1 & \epsilon  & \epsilon  \\[1ex]
\epsilon  & 1 & 1 \\[1ex]
\epsilon  & 1 & 1
\end{array}
\right) M^{2}\;,\quad
{\widetilde{m}}_{d}^{2}\sim \left( 
\begin{array}{ccc}
1 & \epsilon  & \epsilon  \\[1ex]
\epsilon  & 1 & \epsilon  \\[1ex]
\epsilon  & \epsilon  & 1
\end{array}
\right) M^{2}\;.
\eeq
Note, that the zeros of $\widetilde{m}_{l}^{2}$ and $\widetilde{m}_{d}^{2}$
in eq.~(\ref{softmod}) have disappeared since the diagonal part of the matrix 
is not proportional to the identity matrix; the matrix elements are only 
${\cal O}(1)$ (cf.~(\ref{modular})).

To discuss processes involving up quarks, like $t\rightarrow u\gamma$,
$t\rightarrow c\gamma $ or $\Delta M_{D}$, it is convenient to diagonalise 
the up-quark mass matrix. Under this transformation only the form of the 
matrix $\widetilde{m}_{q}^{2}$ is not invariant, and one obtains
\beq
{\widetilde{m}}_{q}^{2}\sim \ \left( 
\begin{array}{ccc}
1 & \epsilon  & \epsilon  \\[1ex]
\epsilon  & 1 & \epsilon  \\[1ex]
\epsilon  & \epsilon  & 1
\end{array}
\right) M^{2}\;.
\eeq

In the $SU(3)^3$ model, one has two scales for the Yukawa couplings, 
$\Lambda_1$ and $\Lambda_2 \gg \Lambda_1$. Consider first the case
$\Lambda_3=\Lambda_2$, which yields the smaller flavour changing soft
breaking terms. For the scalar lepton and quark mass matrices one obtains
from Tab.~\ref{tab:u1f} and eq.~(\ref{modular}), 
\beq
{\widetilde{m}}_{l}^2 \sim {\widetilde{m}}_{e}^2\sim {\widetilde{m}}_{\n}^2
\sim \ \left( 
\begin{array}{lll}
1 & \epsilon^{2} & \epsilon^{5/2} \\ 
\epsilon^{2} & 1 & \epsilon^{1/2} \\ 
\epsilon^{5/2} & \epsilon^{1/2} & 1
\end{array}
\right) \ M^{2}\;,  \label{softmodbis}
\eeq
\beq
{\widetilde{m}}_{q}^{2} \sim \ \left( 
\begin{array}{lll}
1 & \epsilon & \epsilon^{3} \\ 
\epsilon & 1 & \epsilon^{2} \\ 
\epsilon^{3} & \epsilon^{2} & 1
\end{array}
\right) \ M^{2}\;, \quad
{\widetilde{m}}_{u}^{2}\sim {\widetilde{m}}_{d}^{2}\sim \
\left( 
\begin{array}{lll}
1 & \epsilon & \epsilon \\ 
\epsilon & 1 & 0 \\ 
\epsilon & 0 & 1
\end{array}
\right) M^{2}\;.  
\eeq

The scalar mass matrices (\ref{softmodbis}) are again given in the weak 
eigenstate basis. The transition to the mass eigenstate basis is given
by the unitary matrices $U_{e,d}$ and $V_{e,d}$, defined by
$U_{e,d}^{\dagger }h_{e,d}V_{e,d}=h_{e,d}^{D}$, which are now different
for leptons and quarks. To leading order in $\bar{\e}$ one obtains
\beq
V_{e} \sim U_{e}=\left( 
\begin{array}{lll}
1 & a\overline{\epsilon }^{2} & b\overline{\epsilon }^{5/2} \\ 
-a\overline{\epsilon }^{2} & 1 & f\overline{\epsilon }^{1/2} \\ 
-b\overline{\epsilon }^{5/2} & -f\overline{\epsilon }^{1/2}
& 1
\end{array}
\right) \;,  \label{unitaire} 
\eeq
\beq
U_{d} = \left( 
\begin{array}{lll}
1 & a^{\prime }\overline{\epsilon } & b^{\prime }\overline{\epsilon }^{3} \\ 
-a^{\prime }\overline{\epsilon } & 1 & f^{\prime }\overline{\epsilon }^{2}
\\ 
-b^{\prime }\overline{\epsilon }^{3} & -f^{\prime }\overline{\epsilon }^{2}
& 1
\end{array}
\right) \;, \quad V_{d}=\left( 
\begin{array}{ccc}
1 & (c\tilde{a}-s\tilde{b})\overline{\e} & 
(s\tilde{a}+c\tilde{b})\overline{\e} \\[1ex]
-\tilde{a}\overline{\e}\  & c & s \\[1ex]
-\tilde{b}\overline{\e} & -s & c
\end{array}
\right)\; ;
\eeq
here $c=\cos {\varphi }$, $s=\sin {\varphi }$ and $a,\ldots \tilde{b}$ 
depend on the coefficients $\mathcal{O}(1)$ in the Yukawa
matrices. The scalar mass matrices transform as $V_{e,d}^{\dagger }%
\widetilde{m}_{l,q}^{2}V_{e,d}$, $U_{e,d}^{\dagger }\widetilde{m}%
_{e,d}^{2}U_{e,d}$. The form of the matrices $\widetilde{m}_l^2$,
$\widetilde{m}_e^2$ and $\widetilde{m}_u^2$ are invariant under this
transformation. For the other two scalar mass matrices one obtains,
\beq
{\widetilde{m}}_{d}^{2}\sim \ \left( 
\begin{array}{lll}
1 & \overline{\epsilon } & \overline{\epsilon }^2 \\ 
\overline{\epsilon } & 1 & \overline{\epsilon }^{2} \\ 
\overline{\epsilon }^2 & \overline{\epsilon }^{2} & 1
\end{array}
\right) M^{2}\;, \quad
{\widetilde{m}}_{q}^{2}\sim \ \left( 
\begin{array}{ccc}
1 & \overline{\epsilon } & \overline{\epsilon } \\[1ex]
\overline{\epsilon } & 1 & 1 \\[1ex]
\overline{\epsilon } & 1 & 1
\end{array}
\right) M^{2}\; .  \label{softmodbis2}
\end{equation}

The scalar mass matrices in the case $\Lambda_3=\Lambda_1$ are obtained 
by replacing in eqs.~(\ref{softmodbis}) $\bar{\e}$ by $\e$. The change to
a mass eigenstate basis yields essentially again the result 
(\ref{softmodbis2}), with the only difference that now 
$\left(\widetilde{m}_d^2\right)_{13}\sim \overline{\e}$.

For processes involving up quarks, like $t\rightarrow u\gamma,c\gamma$ or
$\Delta M_{D}$ it is convenient to diagonalise the up quark mass matrix. 
Since $u_R$ and $d_R$ belong to the same representation ${\bf Q_R}$ the
resulting mass matrix can be directly obtained from eq.~(\ref{softmodbis2})
by substituting $\widetilde{m}_{d}^{2} \rightarrow \widetilde{m}_{u}^{2}$
and $\bar{\e} \rightarrow \e$.

\subsection{Radiatively induced soft breaking terms}

In models with gravity mediated supersymmetry breaking one usually assumes
at the GUT scale universal scalar masses,
\beq
{\widetilde{m}_{q}}^{2} ={\widetilde{m}_{u}}^{2}={\widetilde{m}_{d}}^{2}={%
\widetilde{m}_{l}}^{2}={\widetilde{m}_{e}}^{2}=M^{2}1\;,  \label{universal}
\eeq
and cubic scalar couplings proportional to the Yukawa couplings,
\beq
\quad A_{e} = h_{e}A\;,\quad A_{\nu }=h_{\nu }A\;,\quad A_{d}=h_{d}A,\quad
A_{u}=h_{u}A \;.
\eeq
Renormalization effects change these matrices significantly at lower scales.
The two main effects are a universal change due to gauge interactions and 
a flavour dependent change due to Yukawa interactions.

For our purposes it is sufficient to treat the effect of Yukawa interactions
in the leading logarithmic approximation.
Integrating the renormalization group equations from the GUT scale, and 
taking the decoupling of heavy fermions at their respective masses $M_{k}$ 
into account, one obtains at scales $\mu \ll M_k$, 
\bea
  (\delta \widetilde{m}_l^2)_{ij} & \simeq & 
  -{\frac{1}{8\pi ^{2}}}%
  (3M^{2}+A^{2}) {h_\nu^\dagger}_{ik} 
  \ln {\frac{\Lambda _{GUT}}{M_{k}}} {h_\nu}_{kj} \; ,  
  \nonumber \\
  (\delta \widetilde{m}_q^2)_{ij} & \simeq &
  -{\frac{1}{8\pi^{2}}} 
  (3M^{2}+A^{2}) {h_u^*}_{ik} 
  \ln {\frac{\Lambda_{GUT}}{M_{k}}} {h_u^T}_{kj} \; , 
  \nonumber \\
  (\delta \widetilde{m}_u^2)_{ij} & \simeq &
  -{\frac{1}{4\pi ^{2}}}%
  (3M^{2}+A^{2}) {h_u^T}_{ik} 
  \ln {\frac{\Lambda _{GUT}}{M_{k}}} {h_u^*}_{kj} \; , 
  \label{softrge} \\
  \delta A_{dij} & \simeq &
  -{\frac{3}{16\pi ^{2}}} A( h_u h_u^{\dagger})_{ik}
  \ln {\frac{ \Lambda_{GUT}} {M_{k}}} {h_d}_{kj} \; ,  
  \nonumber \\
  \delta A_{eij} & \simeq &
  -{\frac{1}{8\pi ^{2}}} A( h_e h_\n^\dagger)_{ik}
  \ln {\frac{\Lambda _{GUT}}{M_{k}}} {h_\nu}_{kj} \; .  
  \nonumber
\end{eqnarray}
Here we have listed only those matrices which have off-diagonal elements
in the mass eigenstate basis of charged leptons and down quarks. This is
not the case for $\delta \widetilde{m}_d^{2}$ and $\delta A_u$.

In the following we shall discuss flavour changing processes to leading order 
in $\e$. We shall not be able to determine factors $\mathcal{O}(1)$. Hence, 
we will neglect terms $\sim \ln {\e^{2}}$, which reflect the splitting between
the heavy neutrino masses, and evaluate $\ln (\Lambda _{GUT}/M_{k})$ for an
average right-handed neutrino mass $\overline{M}=10^{12}$~GeV. This yields
the overall factor $\ln (\Lambda _{GUT}/\overline{M})\sim 10$. The 
flavour changing quark matrices are dominated by the top quark contribution
which gives the factor $\ln (\Lambda_{GUT}/v)\sim 25$.

For $G=SU(5)$, the flavour structure of the scalar mass matrix 
$\delta \widetilde{m}_{l}^{2}$ is identical to the one of the neutrino mass 
matrix, 
\beq
(\delta \widetilde{m}_l^2)_{ij} \sim 
  {\frac{1}{8\pi ^{2}}}(3M^{2}+A^{2})
  \ln {\frac{\Lambda _{GUT}}{\overline{M}}}\ \ \left( 
\begin{array}{ccc}
\epsilon ^{2} & \epsilon  & \epsilon  \\[1ex]
\epsilon  & 1 & 1 \\[1ex]
\epsilon  & 1 & 1
\end{array}
\right) \;.  \label{smix2}
\eeq
For the left-right scalar lepton mass matrix one obtains 
\beq
v_{1} \delta A_{eij} \sim 
{\frac{1}{8\pi ^{2}}}A m_\tau \ln {\frac{\Lambda _{GUT}}{%
\overline{M}}}\ \left( 
\begin{array}{ccc}
\epsilon ^{5} & \epsilon ^{4} & \epsilon ^{4} \\[1ex]
\epsilon ^{2} & \epsilon  & \epsilon  \\[1ex]
\epsilon  & 1 & 1
\end{array}
\right) \;.  \label{smix3}
\eeq
Similarly, one obtains for the three scalar quark mass matrices,  
\beq
(\delta \widetilde{m}_u^2)_{ij} = 
2 (\delta \widetilde{m}_q^2)_{ij} \sim 
{\frac{1}{4\pi ^{2}}}(3M^{2}+A^{2})\ln{\frac{\Lambda _{GUT}}{v}}\ \ \left( 
\begin{array}{ccc}
\epsilon ^{4} & \epsilon ^{3} & \epsilon ^{2} \\[1ex]
\epsilon ^{3} & \epsilon ^{2} & \epsilon  \\[1ex]
\epsilon ^{2} & \epsilon  & 1
\end{array}
\right)\;, 
\eeq
\beq
v_{1} \delta A_{dij} \sim 
{\frac{3}{16\pi ^{2}}} A m_b \ln {\frac{\Lambda_{GUT}}{v}}\ \left( 
\begin{array}{ccc}
\epsilon ^{7} & \epsilon ^{4} & \epsilon ^{2} \\[1ex]
\epsilon ^{6} & \epsilon ^{3} & \epsilon  \\[1ex]
\epsilon ^{5} & \epsilon ^{2} & 1
\end{array}
\right)\;.  
\eeq

For the symmetry group $G=SU(3)^{3}$, the scalar lepton mass matrices read, 
\bea
(\delta \widetilde{m}_l^2)_{ij} &\sim & {\frac{1}{8\pi ^{2}}}%
(3M^{2}+A^{2})\ln {\frac{\Lambda _{GUT}}{\overline{M}}}\ \ \left( 
\begin{array}{ccc}
\epsilon ^{5/2} & \epsilon ^{3/2} & \epsilon ^{5/4} \\[1ex]
\epsilon ^{3/2} & \epsilon^{1/2}  & \epsilon ^{1/4} \\[1ex]
\epsilon ^{5/4} & \epsilon ^{1/4} & 1
\end{array}
\right)\;,  \\
v_{1} \delta A_{eij} & \sim &
{\frac{1}{8\pi ^{2}}}A m_\tau 
\ln {\frac{\Lambda_{GUT}}{\overline{M}}}\ \ \left( 
\begin{array}{ccc}
\epsilon ^{5} & \epsilon ^{4} & \epsilon ^{15/4} \\[1ex]
\epsilon ^{2} & \epsilon  & \epsilon^{3/4}  \\[1ex]
\epsilon ^{5/4} & \epsilon ^{1/4} & 1
\end{array}
\right)\;.
\eea
The corresponding scalar quark mass matrices are given by
\bea
(\delta \widetilde{m}_q^2)_{ij} & \sim & 
{\frac{1}{8\pi ^{2}}}%
(3M^{2}+A^{2})\ln {\frac{\Lambda _{GUT}}{v}}\ \ \left( 
\begin{array}{ccc}
\epsilon ^{3} & \epsilon ^{5/2} & \epsilon ^{3/2} \\[1ex]
\epsilon ^{5/2} & \epsilon ^{2} & \epsilon  \\[1ex]
\epsilon ^{3/2} & \epsilon  & 1
\end{array}
\right)\;, \\
(\delta \widetilde{m}_u^2)_{ij} &\sim &{\frac{1}{4\pi ^{2}}}%
(3M^{2}+A^{2})\ln {\frac{\Lambda _{GUT}}{v}}\ \ \left( 
\begin{array}{ccc}
\epsilon ^{2} & \epsilon  & \epsilon  \\[1ex]
\epsilon  & 1 & 1 \\[1ex]
\epsilon  & 1 & 1
\end{array}
\right)\;,\\ 
v_{1} \delta A_{dij} &\sim &
{\frac{3}{16\pi ^{2}}}A m_b \ln {\frac{\Lambda_{GUT}}{v}}\ \ \left( 
\begin{array}{ccc}
\epsilon ^{5} & \epsilon ^{7/2} & \epsilon ^{3/2} \\[1ex]
\epsilon ^{9/2} & \epsilon ^{3} & \epsilon  \\[1ex]
\epsilon ^{7/2} & \epsilon ^{2} & 1
\end{array}
\right)\;. 
\eea
In order to simplify the comparison with the $SU(5)$ nodel, all mass 
matrices have been expressed in terms of $\epsilon$ (cf.~(\ref{epsilon})).

Gauge interactions affect all scalar masses universally. For the average
scalar quark and lepton masses,
\beq
\overline{M}^2_q = {1\over 9} \mbox{Tr}\left(\widetilde{m}^2_q + 
\widetilde{m}^2_u + \widetilde{m}^2_d\right)\;, \quad
\overline{M}^2_l = {1\over 6} \mbox{Tr}\left(\widetilde{m}^2_l + 
\widetilde{m}^2_e \right)\;, 
\eeq
one obtains at the Fermi scale $\m = v$ \cite{dps96},
\beq\label{renM}
\overline{M}^2_q \simeq M^2 + 7 m^2\;, \quad 
\overline{M}^2_l \simeq M^2 + 0.3 m^2\;, 
\eeq
where $m$ is the universal gaugino mass at the GUT scale. The corresponding
bino, wino and gluino masses are given by \cite{dps96}
\beq\label{renm}
m_{\bi} \simeq 0.4 m\;, \quad m_{\wi} \simeq 0.8 m\;,
\quad m_{\gl} \simeq 3 m\;.
\eeq  

\section{Quark flavour changing processes}

We are now in a position to study specific flavour changing hadronic
processes. We shall compare the four models with symmetry groups
$SU(5)\otimes U(1)_{F}$ and $SU(3)^3\otimes U(1)_{F}$, both with either
modular invariance (MI) or radiatively induced flavour mixing (RI).
 
Given the results of the previous section we can compute the standard model 
predictions as well as the magnitude of the additional supersymmetric 
contributions to leading order in $\e$. We approximate the supersymmetric 
contributions by the gluino exchange terms which dominate for most of the 
parameter space. The flavour changing processes depend on the off-diagonal 
part of the two $6\otimes 6$ scalar mass matrices which are conveniently
written as,
\beq
\frac{\delta{\widetilde{M}_{u,d}}^{2}}{M^{2}}\equiv \left( 
\begin{array}{ll}
\delta^{u,d}_{LL} & \delta^{u,d}_{LR} \\ 
\delta^{u,d}_{RL} & \delta^{u,d}_{RR}
\end{array}
\right)\;. 
\eeq
The electroweak and the supersymmetric contributions have the 
same order of magnitude, since the Fermi scale $v$ and the supersymmetry 
breaking scale $M$ are roughly the same. Hence, it is useful to compare 
directly the powers in $\e$ of the standard model (SM) contributions,
given by the CKM matrix, and the supersymmetric contributions, determined by 
the matrices $\delta_{MN}$ ($M,N= 1,2,3)$, respectively. For the various
processes, the powers of $\e$ are given in Tab.~\ref{tab:fcnc} for the two 
groups $G=SU(5)$ and $G=SU(3)^3$, respectively. 

\begin{table}
\begin{center}
\begin{tabular}{||c||c||cc||ccc||}
\hline \hline
$SU(5)\otimes U(1)_{F}$ & SM & \multicolumn{2}{|c||}{MI} &  & RI &  \\ 
& & $\left(\delta_{LL}\right)^{2}$ & $\left(\delta_{RR}\right)^2$ 
& $\left(\delta_{LL}\right)^2$ & $\left(\delta_{LR}\right)^2$ 
& $\left(\delta_{RL}\right)^2$ \\ 
\hline
$\Delta M_{B_{d}}$ & $\epsilon ^{4}$ & $\epsilon ^{2}$ & $\epsilon ^{2}$ & $%
\epsilon ^{4}$ & $\epsilon ^{4}$ & $\epsilon ^{10}$ \\ 
$\Delta M_{B_{s}},b\rightarrow s\gamma $ & $\epsilon ^{2}$ & $1$ & $\epsilon
^{2}$ & $\epsilon ^{2}$ & $\epsilon ^{2}$ & $\epsilon ^{4}$ \\ 
$\Delta M_{K}$ & $\epsilon ^{4}$ & $\epsilon ^{2}$ & $\epsilon ^{2}$ & $%
\epsilon ^{6}$ & $\epsilon ^{8}$ & $\epsilon ^{12}$ \\ 
$\Delta M_{D}$ & $\epsilon ^{5}$ & $\epsilon ^{2}$ & $\epsilon ^{2}$ & $%
\epsilon ^{6}$ & $\epsilon ^{10}$ & $\epsilon ^{17}$ \\ 
$t\rightarrow u\gamma $ & $\epsilon ^{6}$ & $\epsilon ^{2}$ & $\epsilon ^{4}$
& $\epsilon ^{4}$ & $\epsilon ^{4}$ & $\epsilon ^{12}$ \\ 
$t\rightarrow c\gamma $ & $\epsilon ^{4}$ & $\epsilon ^{2}$ & $\epsilon ^{2}$
& $\epsilon ^{2}$ & $\epsilon ^{2}$ & $\epsilon ^{6}$ \\ \hline
%
\hline 
$SU(3)^{3}\otimes U(1)_{F}$ & SM &  \multicolumn{2}{|c||}{MI} &  & RI &  \\ 
&  & $\left(\delta_{LL}\right)^2$ & $\left(\delta_{RR}\right)^2$ 
& $\left(\delta_{LL}\right)^2$ 
& $\left(\delta_{LR}\right)^2$ & $\left(\delta_{RL}\right)^2$ \\ 
\hline
$\Delta M_{B_{d}}$ & $\epsilon^{3}$ & $\e$ & $\epsilon ^{2}$ & $%
\epsilon ^{3}$ & $\epsilon ^{3}$ & $\epsilon ^{4}$ \\ 
$\Delta M_{B_{s}},b\rightarrow s\gamma $ & $\epsilon ^{2}$ & $1$ & $\epsilon
^{2}$ & $\epsilon ^{2}$ & $\epsilon ^{2}$ & $\epsilon ^{7}$ \\ 
$\Delta M_{K}$ & $\epsilon^{4}$ & $\e $ & $\epsilon ^{2}$ & $%
\epsilon ^{5}$ & $\epsilon ^{7}$ & $\epsilon ^{9}$ \\ 
$\Delta M_{D}$ & $\epsilon ^{5}$ & $\epsilon ^{2}$ & $\epsilon ^{2}$ & $%
\epsilon ^{5}$ & $\epsilon ^{9}$ & $\epsilon ^{13}$ \\ 
$t\rightarrow u\gamma $ & $\epsilon ^{6}$ & $\epsilon ^{2}$ & $\e^4 $
& $\epsilon^3$ & $\epsilon^3$ & $\epsilon^{11}$ \\ 
$t\rightarrow c\gamma $ & $\epsilon ^{4}$ & $1$ & $\epsilon ^{4}$ & $%
\epsilon ^{2}$ & $\epsilon ^{2}$ & $\epsilon ^{6}$ \\ 
\hline \hline
\end{tabular}
\caption[dum]{\textit{Order of magnitude of flavour changing processes for
the groups $G=SU(5)$ and $G=SU(3)^3$ in the standard model (SM) and its 
supersymmetric extensions, with modular invariance (MI) and radiatively 
induced flavour mixing (RI), respectively.}}
\label{tab:fcnc}
\end{center}
\end{table}

For completeness we have also listed flavour changing t-decays 
and $\D M_D$ where the
supersymmetric contributions dominate over the standard model terms. At
a future linear $e^+e^-$ collider the transition $t\rightarrow c\g$ may be
observable.

As one reads off the table, the models with modular invariance appear
to yield predictions larger than the standard model ones. Since these are
in agreement with data, models with modular invaariance are clearly in danger 
of being ruled out. A detailed discussion of several processes will be given 
in the following subsections.

In this connection it is important to recall the renormalization of the
diagonal part of the scalar quark mass matrix which can be quite large 
for large gaugino masses (cf.~(\ref{renM})), as emphasized by Choudhury 
et al. \cite{cek95}.

\subsection{The $K-\overline{K}$ system}

A neutral meson formed by a 
heavy quark $Q$ and a light quark $q$, i.e. $M=(Q\overline{q})$, can mix with 
the corresponding anti-meson $\overline{M}=(\overline{Q}q)$.
The strength of the mixing is determined by the matrix element of the
effective hamiltonian which changes the $Q$-number by two units, 
\beq
M_{12}={\frac{1}{2m_{M}}}\langle \overline{M}|\mathcal{H}_{eff}^{\Delta
Q=2}|M\rangle \; .
\eeq
The mass difference between the mass eigenstates is given by the real part, 
$\Delta M_{M}\simeq 2$Re$M_{12}$. The imaginary part Im$M_{12}$ can be
measured in CP-violating decays. For the $K^{0}$ meson one obtains 
within the standard model \cite{bf98,bl99}, 
\begin{equation}
\Delta M_{K}={\frac{G_{F}^2}{6\pi ^{2}}}\eta 
_{K}m_{K}B_{K}f_{K}^{2}m_{W}^{2}S_{0}(x_{c})|V_{cd}^{*}V_{cs}|^{2}\;.
\label{dmk1}
\end{equation}
Here $G_{F}$ is the Fermi constant, $\eta _{K}=\mathcal{O}(1)$ is a QCD
correction factor, $f_{K}$ is the K-decay constant and $B_{K}$ reflects the
matrix element of the 4-fermion operator. For $x_{c}=m_{c}^{2}/m_{W}^{2}\ll 1
$ one has $S_{0}(x_{x})\simeq 10x_{c}/4$. In (\ref{dmk1}) we have neglected
the top-quark contributions which turn out not to contribute to leading order 
in $\epsilon $. 

For comparison with supersymmetric contributions it is convenient to
replace $G_{F}$ and $m_{W}$ by the Higgs vacuum expectation value $v\simeq
174$~GeV and by the SU(2) fine structure constant $\alpha _{2}$
respectively, which yields
\begin{equation}
\Delta M_{K}={\frac{\alpha _{2}}{24\pi }}{\frac{1}{v^{2}}}%
|V_{cd}^{*}V_{cs}|^{2}m_{K}B_{K}f_{K}^{2}\eta _{K}S_{0}(x_{c})\;.
\label{dmk}
\end{equation}
The CP violation parameter is given by\cite{bf98,bl99} 
\begin{eqnarray}
  \varepsilon _{K} & = & 
  {\frac{e^{i\pi /4}}{\sqrt{2}\Delta M_{K}}}\mbox{Im}M_{12} \\
  & = &
  {\frac{\alpha _{2}}{24\pi }}{\frac{1}{v^{2}}}
  \mbox{Im}(V_{td}^{*}V_{ts}) 
  \left[ \mbox{Re}(V_{cd}^{*} V_{cs})
    \left( \eta _{1}S_{0}(x_{c})-\eta_{3}S_{0}(x_{c},x_{t}) \right) \right. 
  \nonumber \\
  & & 
  \hspace{3cm}\left. - 
    \mbox{Re}(V_{td}^{*}V_{ts}) \eta _{2}S_{0}(x_{t}) \right] 
  {\frac{m_{K}B_{K}f_{K}^{2}}{\sqrt{2}\Delta M_{K}}}e^{i\pi /4}\; .  
  \label{cpk}
\end{eqnarray}
Again, $\eta_1$, $\eta_2$, $\eta_3$ are the QCD correction factors.

From eqs.~(\ref{dmk}) and (\ref{cpk}) one easily obtains the leading order 
in $\e$ for $\Delta M_{K}$ and $\varepsilon_{K}$. Using 
$S_{0}(x_{c})\sim S_{0}(x_{c},x_{t})\sim \epsilon ^{2}$, $S_{0}(x_{t})\sim
\epsilon ^{0}$ and the CKM matrices (\ref{ckm5}) and (\ref{ckm3}) for the
symmetries $G=SU(5)$ and $G=SU(3)^{3}$, respectively, one finds in both cases 
\beq
\Delta M_{K}\sim \epsilon ^{4}\;,\quad \varepsilon _{K}\sim \epsilon ^{2}\;.
\end{equation}

This has to be compared with the gluino contributions, for which the
effective hamiltonian has been studied in detail by Gabbiani et al.
\cite{ggm96}. To estimate the order of magnitude we consider the gluino 
contribution for the special choice of masses $m_{\gl}=M$. From \cite{ggm96}
one obtains, 
\bea
 \Delta M_{K} & \simeq &
 {\frac{\alpha_s^2}{324}}{\frac{1}{M^{2}}}
 m_{K}B_{K}f_{K}^{2} \left\{ 
   (\delta^d_{LL})_{12}^2 + (\delta^d_{RR})^2_{12}) 
 \right. 
 \nonumber\\
 & &\left. 
   - \frac{2}{5} \left[ 6 (\delta^d_{LL})_{12} (\delta^d_{RR})_{12}
   + 7 (\delta^d_{LR})_{12} (\delta^d_{RL})_{12} \right] 
 \right. \nonumber\\
 & &\left. 
   + 
   {\frac{1}{5}}\left({\frac{m_{K}}{m_{d}+m_{s}}}\right)^{2}
   \left[ -100 (\delta^d_{LL})_{12} (\delta^d_{RR})_{12} \right. \right.  
 \nonumber  \\
 & & \left.\left. \hspace{3cm} 
 + 33 \left( (\delta^d_{LR})_{12}^2 + (\delta^d_{RL})_{12}^2 \right)
   - 24 (\delta^d_{LR})_{12} (\delta^d_{RL})_{12} \right] 
 \right\} \; '  
 \label{sdmk}
\eea
Here we have used $f_{6}(1)=1/20$ and $\tilde{f}_{6}(1)=-1/30$, with 
\bea
f_{6}(x) &=&\frac{17-9x-9x^{2}+x^{3}+6(1+3x)\ln x}{6(1-x)^{5}}\;, \\
\tilde{f}_{6}(x) &=&\frac{1+9x-9x^{2}-x^{3}+6x(1+x)\ln x}{3(1-x)^{5}}\;,
\eea
where $x = m_{\gl}^2/M^2$.
The prefactors in eq.~(\ref{sdmk}) and in the standard model expression are
of the same order of magnitude, since $M\sim v$ and $\alpha _{2}\sim
a_{s}^{2}$. Hence, the relative magnitude of the two contributions is
directly given by the powers in $\e$ of $|V_{cd}^{*}V_{cs}|^{2}$ and
the various $\delta _{IJ}^{d}$, $I,J=L,R$, respectively. From 
eq.~(\ref{sdmk}) and the results given in Sec.~2.2 one obtains for the
models with modular invariance, 
\beq
\begin{array}{lcl}
\Delta M_{K}^{MI} & = & \left\{ 
\begin{array}{lcl}
\epsilon ^{2} & {\rm for} & G = SU(5) \\ 
\epsilon & {\rm for} & G = SU(3)^{3} \; .
\end{array}
\right. 
\end{array}
\eeq
Hence, for both symmetry groups, the supersymmetric contribution appears
to be several orders of magnitude larger than the standard model one,
which is known to be in agreement with observation.

Does this mean that models with modular invariance are excluded? It may be
possible to avoid this conlusion by a different assignment of $U(1)_F$
charges \cite{cek95}. Yet such a choice of charges has to be consistent
with unification and a large $\n_\m - \n_\t$ mixing angle. 

However, models with modular invariance can be in agreement with the observed 
$K^0-\overline{K}^0$ mixing if the gaugino masses are sufficiently large
\cite{cek95}. From eq.~(\ref{renM}) one reads off that 
$M^2/\overline{M}_q^2 < \e$ for $m > \sqrt{2} M$. This is sufficient to
suppress the $s-d$ mixing below the standard model prediction.
This requirement leads to an interesting prediction. For $m > \sqrt{2} M$
the scalar quark masses at the Fermi scale are dominated by the radiatively
induced gluino contribution. From eqs.~(\ref{renM}) and (\ref{renm}) one then obtains
for the ratio of average scalar quark and gluino masses,
\beq
{\overline{M}^2_q\over m_{\gl}^2} \simeq {7\over 9}\;.
\eeq

For radiatively induced flavour mixing the supersymmetric contribution
to $\Delta M_K$ is suppressed with respect to the standard model contribution
(cf. Tab. \ref{tab:fcnc}). Hence, no strong constraint on 
$\overline{M}^2_q$ can be derived.
  
\subsection{The $B-\bar{B}$ system}

\begin{figure}[!b]
  \centering 
  \includegraphics[scale=0.6]{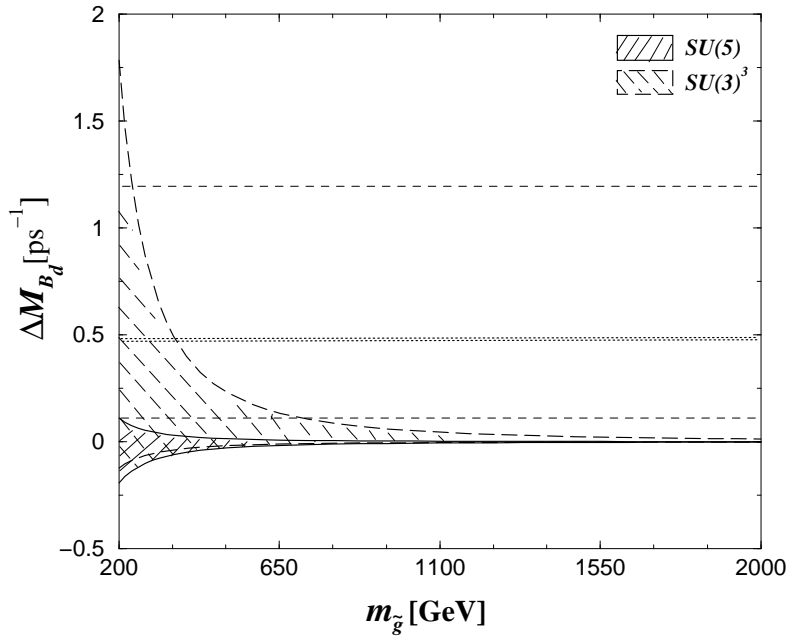}
\caption{{\it Gluino contributions to $\Delta M_{B^0}$ for the radiatively
induced models. The horizontal band represents the measured mass difference, 
and the horizontal dashed lines denote upper and lower bound of the SM 
prediction for $\sqrt{f_B^2 B_B}= 215 \mbox{MeV}$.}}
\label{fig:bb}
\end{figure}

The case of the $B-\bar{B}$ system is very similar to the $K-\bar{K}$
system. The standard model contribution to $\Delta M_{B^0}$ is given by 
\beq
\Delta M_{B^0}={\alpha_2\over 24\pi}{1\over v^2}|V_{td}^{*}V_{tb}|^{2}
m_{B}B_{B}f_{B}^{2}\eta_B S_{0}(x_{t})\;,
\eeq
with $\eta_{B}=0.55$. The supersymmetric contribution due to gluino exchange
is dominated by the $\left(\delta_{LL}^d\right)^2$ part, since the 
$\left(\delta_{RL}^d\right)^2$ and $\left(\delta_{LR}^d\right)^2$
terms are suppressed by $m_b^2$. One then has \cite{ggm96},
\beq
\Delta M_{B^0}^{\gl} = \frac{\alpha_s^2}{54 M^2} m_B f_B^2 
 (\delta^d_{LL})_{13}^2 \left(4 x f_6(x) + 11 \tilde{f}_6(x)\right) \; ,
\label{eq:bdbd}
\eeq
where $x=m_{\gl}^2/M^2$.
In Fig.~\ref{fig:bb} the supersymmetric contributions for the two models with
radiatively induced flavour mixing are compared with the observed mass
difference \cite{pdg98} and the standard model prediction \cite{bf98}. In 
order to determine the
uncertainty of the supersymmetric contribution one has to vary the various 
mass parameters in a range consistent with present experimental limits. 
We have chosen $m_{\tilde{g}}=200\ldots 2000$~GeV and $M > {m_{\tilde{g}}}/2$.
Due to the special properties of the functions $f_6$ and $\tilde{f}_6$ this 
is equivalent to the entire range of $x$ in (\ref{eq:bdbd}). To estimate the 
uncertainty due to the unknown coefficients ${\mathcal{O}}(1)$ we have
multiplied the upper (lower) bound obtained from eq.~(\ref{eq:bdbd}) by
5 (1/5).

Note, that because of the large uncertainty of $V_{td}$ the agreement between 
standard model prediction and observation does not impose a significant 
constraint on the masses of scalar quarks and gluino.
 
\subsection{$b\rightarrow s\gamma $}

The radiative $B$ meson decay is governed by the effective hamiltonian, 
\beq
{\cal H}_{\rm eff}({B \rightarrow X_s\,\g})= - \frac{4 G_F}{ \sqrt{2}}
 V^*_{ts}V_{tb} 
 \sum_{i=1}^{8}{\cal C}_i(\m)\ {\cal O}_i(\m)\;.\label{eq:hebqg}
\eeq
The dominant operators are ${\cal O}_{7}$ and ${\cal O}_{7}^{\prime}$,
\bea
{\cal O}_{7} &=&\frac{e}{16\pi ^{2}}\,m_b\,\left(\bar{s_L}\,\sigma
_{\mu \nu }\, b_R\right) \,F^{\m\n}\;,  \label{eq:o7} \\
{\cal O}_{7}^{\prime} &=&\frac{e}{16\pi ^{2}}\,m_b\,\left( \bar{s_R}%
\,\sigma _{\mu \nu }\,b_L\right) \,F^{\mu \nu }\;,  \label{eq:o7p}
\eea
where $F^{\m\n}$ is the electromagnetic field strength tensor. Note that 
terms ${\cal O}(m_s)$ are neglected. The other operators
contribute mostly through mixing and effect the evolution of the Wilson 
coefficients from $\mu\sim m_W$ to $\mu \sim m_b$.

\begin{figure}[!h]
  \centering 
  \includegraphics[scale=0.6]{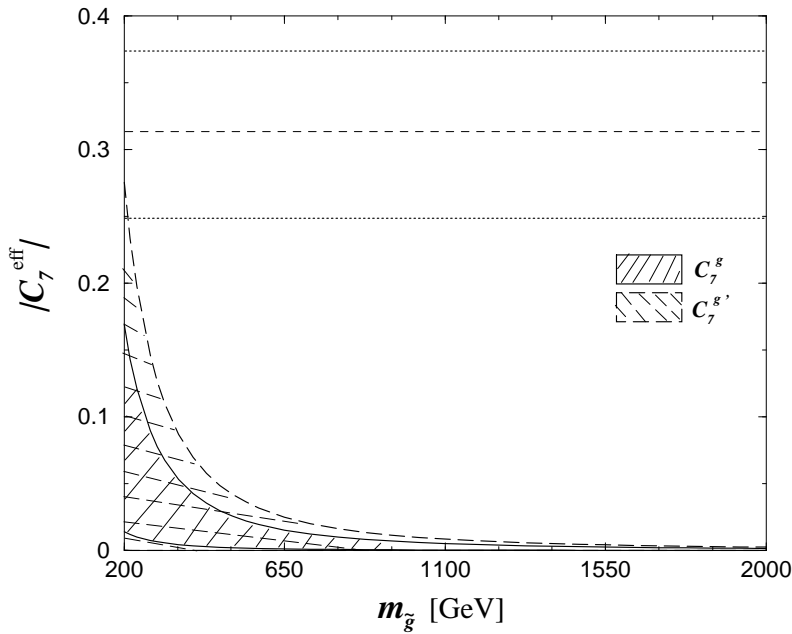}
\caption{{\it Gluino contributions to $C_7^{\mathrm{eff}}$ in the two
models with radiatively induced flavour mixing. The horizontal dotted lines 
denote the experimental bounds, the dashed line represents the SM
contribution.}}
\label{fig:bsg}
\end{figure}

The standard model contribution only affects the Wilson coefficient 
${\cal C}_7$, whereas the gluino exchange contributes to both 
${\cal C}_7^g$ and ${\cal C}_7^{g \prime}$. Hence the branching ratio 
$\mathrm{BR}(B \rightarrow X_s\gamma)$ is determined by the parameter
\beq
\left|{\cal C}_7^{eff}\right|^2 
= \left|{\cal C}_7^{\mathrm{SM}} + {\cal C}_7^g\right|^2 
+ \left|{\cal C}_7^{g \prime} \right|^2\;.
\eeq

The SM contribution has been calculated up to next-to-leading order accuracy,
yielding the result 
$({\cal C}_7^{\mathrm{SM}})^{\mathrm{NLO}}(\m=m_b)=-0.305$. The gluino 
contributions are given in \cite{cmw96},
\bea
{\cal C}_7^g &=&\frac{\sqrt{2}\pi\alpha_s}{3G_F}
   \frac{1}{V_{ts}^*V_{tb}}\frac{N_c^2-1}{2N_c}\frac{1}{M^2}
   \left[(\delta^d_{RL})_{23} \frac{m_{\tilde{g}}}{m_b} F_3(x) 
   + (\delta^d_{LL})_{23} F_4 (x)\right] \;, \\
{\cal C}_7^{g \prime} &=&
  \frac{\sqrt{2}\pi\alpha_s}{3G_F}\frac{1}{V_{ts}^*V_{tb}}
  \frac{N_c^2-1}{2N_c}\frac{1}{M^2}
  \left[ (\delta^d_{LR})_{23} \frac{m_{\tilde{g}}}{m_b} F_3(x) + 
    (\delta_{RR}^d)_{23} F_4(x) \right] \; ,
\eea
where $x=m_{\tilde{g}}^2/M^2$ and
\bea
F_{3}(x) &=&\frac{1+4x-5x^{2}+2x(2+x)\ln x}{2(1-x)^{4}}\;, \\
F_{4}(x) &=&\frac{1-9x-9x^{2}+17x^{3}-6x^{2}(3+x)\ln x}{12(1-x)^{5}}\;.
\eea

The experimental bounds on the branching ratio,
$2.0\cdot 10^{-4}\leq \mathrm{BR}^{\mathrm{exp}}({B \rightarrow X_s\ \g})
\leq 4.5\cdot 10^{-4}$\cite{cle99}
yield the constraint 
$0.249 < \left|{\cal C}_7^{\mathrm{eff}}({B \rightarrow X_s\ \g}) \right| 
< 0.374$ at leading-log accuracy.
In Fig.~\ref{fig:bsg} 
this is compared with the SM prediction and the absolute value
of the gluino contributions ${\cal C}_7^g$ and ${\cal C}_7^{g ^\prime}$.
The interference with the SM contribution can not be discussed since
the quantities $\delta_{LL}^d,\ldots \delta_{RL}^d$ are only known up
to terms ${\cal O}(1)$. From Tab.~\ref{tab:fcnc} it is clear that the two 
symmetry groups $G=SU(5)$ and $G=SU(3)^3$ yield similar results. 
As Fig.~\ref{fig:bsg} illustrates, the gluino contributions can be 
significant for gluino masses below 500~GeV.

\subsection{Electric dipole moments}

For comparison and completeness we also recall the theoretical predictions
for the electric dipole moments in supersymmetric extensions of the
standard model \cite{bw83}. A more recent thorough discussion has
been given in \cite{bs91} for neutron and electron. One obtains for
$m_{\gl}=M$ and $m_{\bi}=M$, respectively,
\bea
{d_n\over e} &\simeq & 2\cdot 10^{-23} \left({100\mbox{GeV}\over M}\right)^3
{A\over 100\mbox{GeV}} \sin{\alpha}\;, \\
{d_e\over e} &\simeq & 1\cdot 10^{-25} \left({100\mbox{GeV}\over M}\right)^3
{A\over 100\mbox{GeV}} \sin{\gamma}\;, 
\eea
where $\a$ and $\g$ are CP violating angles.
As an example, for $M=A=100$~GeV the predictions exceed the experimental
upper bounds $d_n^{exp}/e < 6.3\cdot 10^{-26}$~cm \cite{ha99} and
$d_e^{exp}/e < 4.3\cdot 10^{-27}$~cm \cite{co94} by almost two orders
of magnitude. Hence, either the CP violating angles are small due to some
approximate symmetry or scalar and gaugino masses are in a range which
makes the observation of lepton flavour changing transitions also difficult.

\section{Lepton flavour changing processes}

\begin{figure}[!h]
  \centering 
  \includegraphics[scale=0.6]{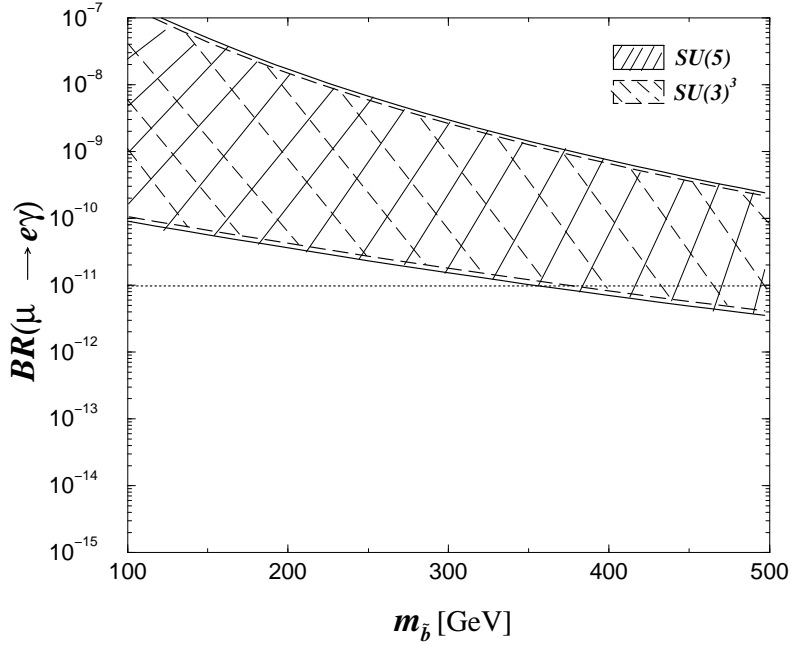}
  \vspace*{1cm}\\
  \includegraphics[scale=0.6]{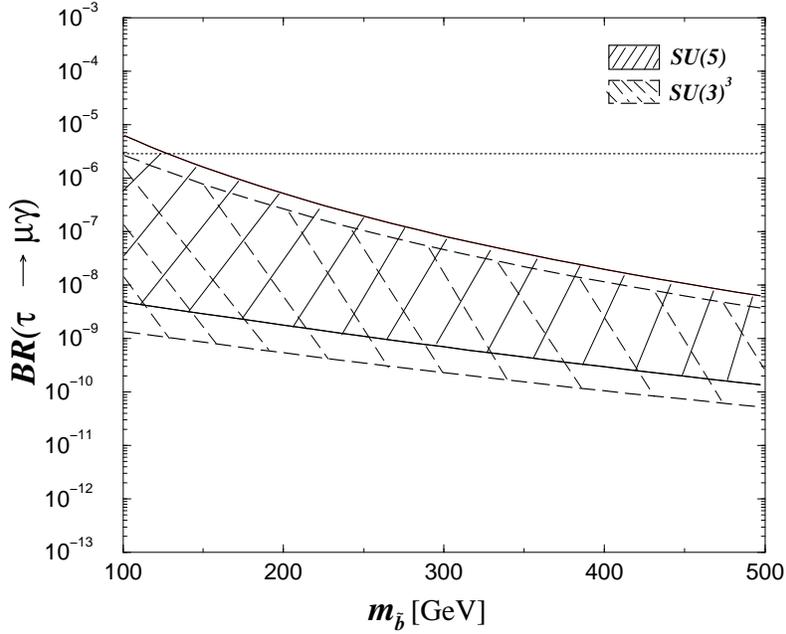}
\caption{{\it Predicted range for $\mathrm{BR}(\mu \rightarrow e \gamma)$ and 
$\mathrm{BR}(\tau \rightarrow \mu \gamma)$ as function of the bino mass for two
models with modular invariance based on the symmetry groups
$G=SU(5)$ and $G=SU(3)^3$, respectively. The straight lines represent the
experimental bounds \cite{bbb}.}}
\label{fig:lfmi}
\end{figure}

\begin{figure}[!h]
  \centering 
  \includegraphics[scale=0.6]{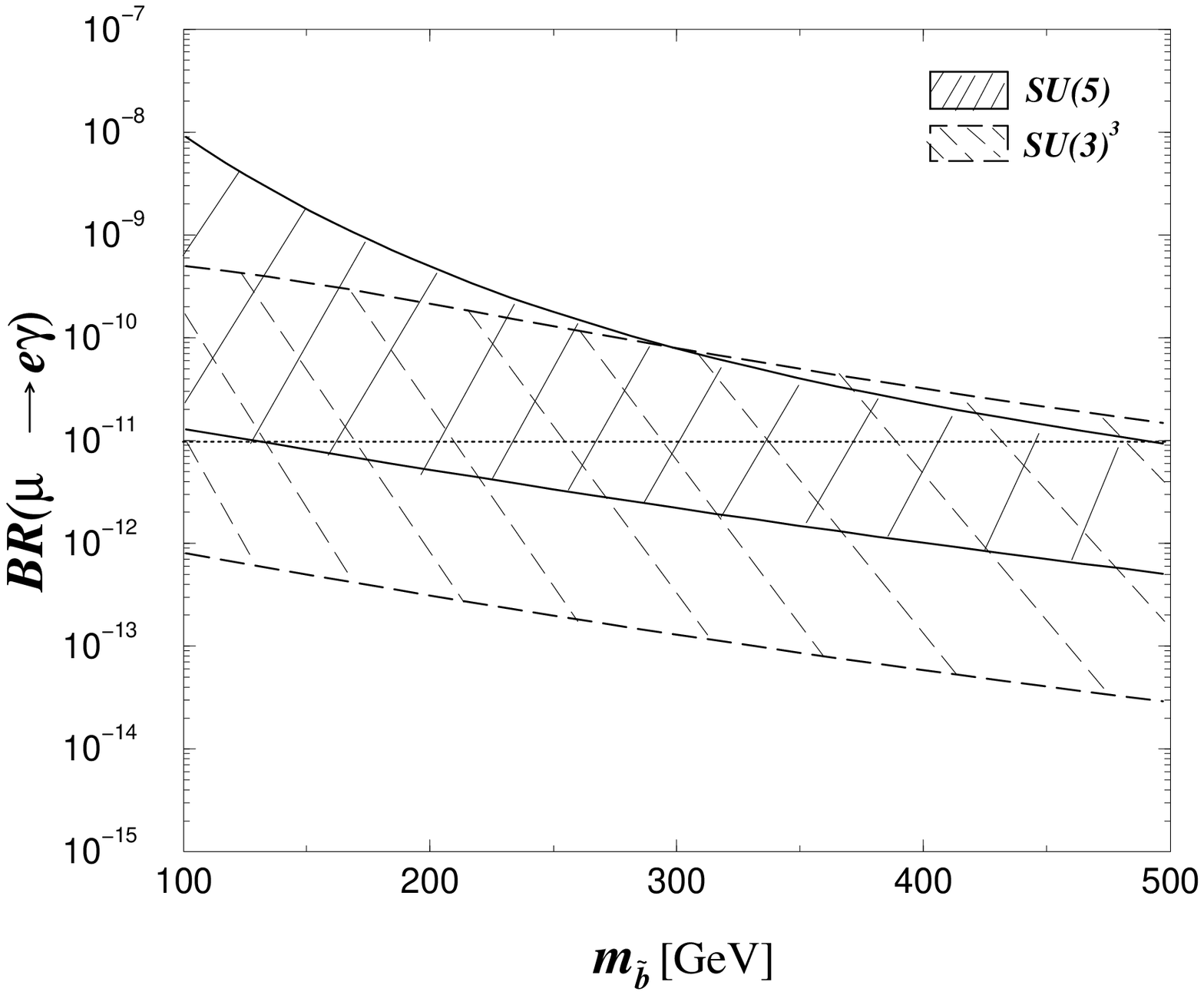}
  \vspace*{1cm}\\
  \includegraphics[scale=0.6]{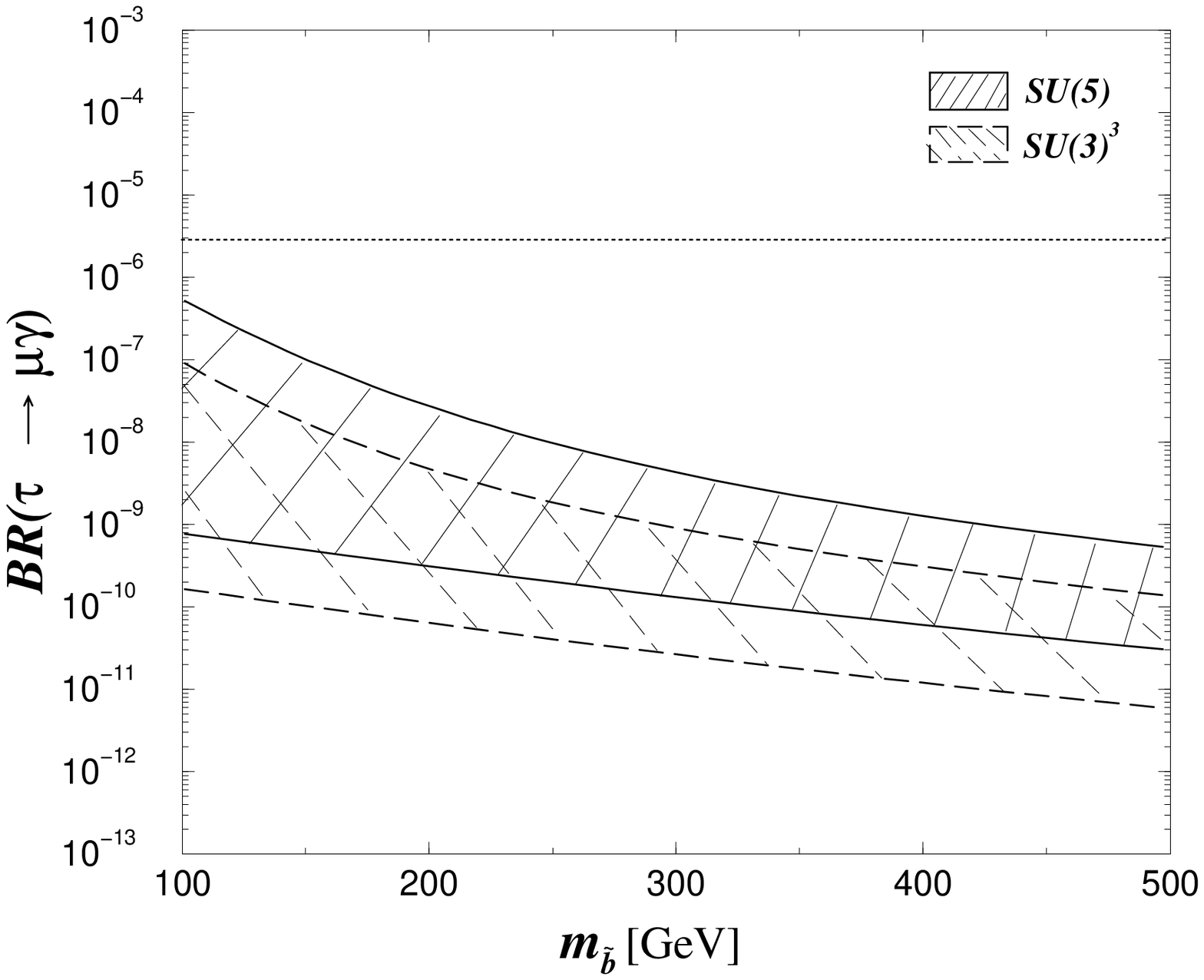}
\caption{{\it Predicted range for $\mathrm{BR}(\mu \rightarrow e \gamma)$ and 
$\mathrm{BR}(\tau \rightarrow \mu \gamma)$ as function of the bino mass for two
models with radiatively induced flavour mixing based on the symmetry groups
$G=SU(5)$ and $G=SU(3)^3$, respectively. The straight lines represent the
experimental bounds \cite{bbb}.}}
\label{fig:lfri}
\end{figure}

Particularly interesting processes 
are lepton flavour changing transitions, which have
already been discussed for radiatively induced flavour mixing and $G=SU(5)$ 
in ref.~\cite{bdv99}. The enhancement of scalar lepton masses at the
Fermi scale due to renormalization is much smaller than for scalar quark
masses (cf.~(\ref{renM})). Hence, lepton flavour changing processes
are generically larger than quark flavour changing processes \cite{dps96}. 
For large Yukawa couplings, branching ratios comparable to present 
experimental limits are predicted.

Given the Yukawa matrices and the scalar mass matrices it is straightforward
to calculate the rates for radiative transitions. The transition amplitude 
$\m \rightarrow e\g$ has the form 
\beq
 \mathcal{M}_{\m} = i e \bar{u}_{e}(p-q) \sigma_{\mu\nu} q^\nu
 \left[ (A_L)_{12} P_L + (A_R)_{12} P_R \right] 
 u_\mu(p) \; ,  
 \label{transamp}
\eeq
where $P_{L}$ and $P_{R}$ are the projectors on states with left- and
right-handed chirality, respectively. The corresponding branching ratio is
given by 
\beq
 \mathrm{BR}(\mu \rightarrow e \gamma ) = 
 384\pi ^{3}\alpha {\frac{v^{4}}{m_\mu^{2}}}%
 (|(A_L)_{12}|^{2}+|(A_R)_{12}|^{2})\;,  \label{brmu}
\eeq
where $v=(8G_{F}^{2})^{-1/4}\simeq 174$~GeV is the Higgs vacuum expectation
value. Analogously, using $\G_\t \simeq 5 (m_\t/m_\m)^5 \G_\m$, one obtains 
for the process $\t \rightarrow \m \g$,
\beq
  \mathrm{BR}(\t \rightarrow \m\g) = {384\p^3\over 5} \a {v^4\over m_\t^2}
  (|(A_L)_{23}|^2 + |(A_R)_{23}|^2)\;.
\eeq
The amplitudes $A_{L12}$...$A_{R23}$ have been given explicitly in 
ref.~\cite{bdv99}.

The results for $\mu \rightarrow e\gamma $ and  $\tau \rightarrow
\mu \gamma $ are plotted in Figs.~\ref{fig:lfmi} and 
\ref{fig:lfri}. In order to determine the 
uncertainty of the theoretical predictions we again vary the supersymmetry 
breaking mass parameters in a range consistent with experimental limits. 
Following \cite{bdv99} we choose for gaugino masses and the average scalar
mass $m_{\bi}=m_{\wi}=100\ldots 500$~GeV, $M=100\ldots 500$~GeV, 
$A=0\ldots M$, $A+\mu \tan {\beta }=0\ldots M$. Since we know the transition 
amplitude only up to a factor $\mathcal{O}(1)$, we neglect neutralino and 
chargino mixings. To estimate these uncertainties we increase the upper bound 
by a factor of $5$ and decrease the lower bound by a factor $1/5$. 

The branching ratios for $\m \rightarrow e\g$ and for $\t \rightarrow \m \g$
are plotted in Figs.~\ref{fig:lfmi} for the models with modular invarance. 
For $G=SU(5)$ the range for $BR(\m \rightarrow e\g)$ agrees with the result
in \cite{bdv99}. For $\t \rightarrow \m \g$ larger branching ratios are
obtained since, contrary to \cite{bdv99}, the most general form of
$\widetilde{m}_l^2$ has been assumed in eq.~(\ref{general}). Consistency
with the experimental upper bound on $\mathrm{BR}(\m \rightarrow e\g)$ yields
a lower bound on the bino mass, $m_{\bi} > 350$~GeV. For 
$\t \rightarrow \m \g$ the predicted branching ratio lies below the present 
experimental bound.

For the models with radiatively induced flavour mixing and large Yukawa
couplings, i.e. $\tan{\b} \sim 1/\e$, both branching ratios are shown in
Figs.~\ref{fig:lfri}. The results for $G=SU(5)$ are identical with the
ones obtained in \cite{bdv99}. The difference between the two cases
$G=SU(5)$ and $G=SU(3)^3$ illustrates the dependence on the pattern of
family charges and the size of the flavour mixing parameter $\e$. 
The rates are comparable to the present experimental upper bound. 
No model independent lower bound on the bino mass can be obtained. It is
very interesting that a branching ratio above $10^{-14}$ is predicted for 
most of the parameter space. This sensitivity is the goal of the recently
approved experiment at PSI \cite{ba99}.

Note, that we have assumed large Yukawa couplings for down quarks and charged
leptons, i.e.  $\tan{\b} \sim 1/\e$. For small Yukawa couplings, i.e. 
$\tan{\b} = {\cal O}(1)$, the branching ratios are smaller by roughly
four orders of magnitude \cite{bdv99}. For part of the parameter space a
branching ratio  $BR(\m \rightarrow e\g) > 10^{-14}$ is predicted also in 
this case.

$\mu-e$ conversion provides also a test of models with radiatively induced 
flavour mixing and large Yukawa couplings. Indeed, assuming that the 
on-shell electromagnetic form factors ($q^2=0$)  dominate the $\mu-e$ 
conversion processes, one obtains \cite{wf59}
\beq
R = \frac{\s (\m^- T_i \rightarrow e^- T_i)}
         {\s(\m^- T_i \rightarrow \mbox{capture})} 
\simeq 5\cdot 10^{-3} BR(\mu \rightarrow e \gamma)\; .
\eeq
The present experimental upper bound is $R < 1.7\cdot 10^{-12}$ 
\cite{sindrum98}. In the near future, a new round at 
SINDRUM-II is expected to improve the sensitivity by about one order of 
magnitude \cite{sindrum98}, and the MECO collaboration aims at a sensitivity 
for $R$ below $10^{-16}$ \cite{meco97}. 

\section{Conclusions}

We have considered flavour changing processes for quarks and leptons.
Motivated by the present hints pointing beyond the standard model, the
unification of gauge couplings and the possible smallness of neutrinos
masses, we have performed our analysis within the framework of supersymmetric
unified theories. In addition we have assumed a $U(1)_F$ family symmetry
which can account for the observed hierarchies of quark and lepton
masses and, in particular, a large $\n_\m - \n_\t$ mixing angle. Further,
two patterns of supersymmetry breaking have been considered, models with
modular invariance and the standard scenario of universal soft breaking 
terms at the GUT scale.

The models with modular invariance are only consistent with 
$K^0-\overline{K^0}$ mixing, $B^0-\overline{B^0}$ mixing, 
$b \rightarrow s \g$, etc., if all flavour changing transitions are 
universally suppressed by the large renormalization effects for
all scalar quark masses caused by gauge interactions.
This implies a fixed ratio of scalar quark masses and the gluino mass at 
the Fermi scale, $\overline{M}_q^2/m_{\gl}^2 \simeq 7/9$. In addition,
the experimental upper bound on $\mathrm{BR}(\m \rightarrow e\g)$ yields a
lower bound on the bino mass, $m_{\bi} > 350$~GeV.

For the models with radiatively induced flavour mixing and large Yukawa
couplings, i.e. $\tan{\b} \sim 1/\e$, no constraints on scalar masses
and gaugino masses can be derived. The predicted branching ratios
are comparable to the present experimental upper bound on 
$\mathrm{BR}(\m \rightarrow e\g)$ and an improvement of the experimental 
sensitivity down to $10^{-14}$ is predicted to yield a positive signal.
However, for small Yukawa couplings, i.e. $\tan{\b} = {\cal O}(1)$, the 
branching ratios are smaller by roughly four orders of magnitude.

In summary, the interplay of large $\n_\m - \n_\t$ mixing, supersymmetry
and $b-t-\t$ unification,  i.e. large $\tan{\b}$, lead to the prediction
of a branching ratio $BR(\m \rightarrow e\g) > 10^{-14}$. Hence, the discovery 
of the transition $\m \rightarrow e\g$ may provide the first hint for
supersymmetry before the start of LHC.\\

\noindent
{\bf Acknowledgements}\\ \\
\noindent 
One of us (W.B.) would like to thank S.~Pokorski for clarifying discussions.
L.~T.~Handoko thanks the Alexander von Humboldt-Stiftung for support.

\end{document}